\documentstyle[numreferences]{kapproc}
\newcommand{\drm}{\rm d}

\newcommand{\Le}{{\cal L}}

\newcommand{\bb}{\begin{equation}}
\newcommand{\ee}{\end{equation}}
\newcommand{\bega}{\begin{eqnarray}}
\newcommand{\ega}{\end{eqnarray}}
\newcommand{\begae}{\begin{eqnarray*}}
\newcommand{\egae}{\end{eqnarray*}}

\newcommand{\ga}{\gamma}

\newcommand{\age}{\dagger}

\newcommand{\sig}{\sigma}

\newcommand{\longr}{\longrightarrow}
\newcommand{\Lef}{\Leftrightarrow}

\newcommand{\vare}{p^{\rm o}}

\newcommand{\h}{\hspace*{0.5 cm}}

\newcommand{\wide}{\widehat}
\newcommand{\ov}{\overline}
\newcommand{\be}{\beta}
\newcommand{\bt}{\beta}

\newcommand{\pa}{\partial}

\newcommand{\nablabf}{\mbox{\boldmath $\nabla$}}
\newcommand{\Vbf}{\mbox{\boldmath $V$}}
\newcommand{\Abf}{\mbox{\boldmath $A$}}

\newcommand{\sbf}{\mbox{\boldmath $s$}}

\newcommand{\gabf}{\mbox{\boldmath $\gamma$}}
\newcommand{\etbf}{\mbox{\boldmath $\eta$}}

\newcommand{\vbf}{\mbox{\boldmath $v$}}
\newcommand{\abf}{\mbox{\boldmath $a$}}
\newcommand{\bbf}{\mbox{\boldmath $b$}}
\newcommand{\albf}{\mbox{\boldmath $\alpha$}}
\newcommand{\wbf}{\mbox{\boldmath $w$}}
\newcommand{\xbf}{\mbox{\boldmath $x$}}

\newcommand{\rbf}{\mbox{\boldmath $r$}}
\newcommand{\jbf}{\mbox{\boldmath $j$}}
\newcommand{\imp}{\mbox{\boldmath $p$}}

\newcommand{\psib}{\bar{\psi}}

\newcommand{\po}{{\wide {p}}}

\newcommand{\xibf}{\mbox{\boldmath $\xi$}}
\newcommand{\Xbf}{\mbox{\boldmath $X$}}

\newcommand{\ug}{\; = \;}

\newcommand{\bi}{\bibitem}

\runningtitle{KINEMATICS AND HYDRODYNAMICS OF SPINNING PARTICLES}
\begin{opening}
\title{About Kinematics and Hydrodynamics of Spinning\\ 
Particles: Some Simple Considerations\thanks{Work partially supported 
by UNAM, by FAPESP, CNPq, and by INFN, MURST, CNR.}}
\author{Erasmo Recami}
\institute{Facolt\`a di Ingegneria, Universit\`a Statale di Bergamo,
24044--Dalmine (BG),\\ Italy; \ 
INFN--Sezione di Milano, Milan, Italy: \ {\rm Recami@mi.infn.it ;} \ \ and\\
Dept. of Applied Math., State University at Campinas, Campinas, S.P.,\\ 
Brazil: \ {\rm Recami@turing.unicamp.br .}}
\author{Giovanni Salesi}
\institute{Dipartimento di Fisica, Universit\`a  Statale di Catania,
95129--Catania, Italy;\\ 
and \ INFN--Sezione di Catania, Catania, Italy: \ 
{\rm Salesi@ct.infn.it .}}

\end{opening}\begin{document}
\begin{center}Submitted 25 November 1995\end{center}
\hyphenation{}
\begin{abstract} \ In the first part (Sections 1 and 2) of this paper 
 ---starting from the 
Pauli current, in the ordinary tensorial language---  we obtain the 
decomposition of the non-relativistic field velocity into two orthogonal
parts: \ (i) the ``classical'' part, that is, the velocity $\vec{\wbf} = 
\vec{\imp} / m$
{\it of} the center-of-mass (CM), and \ (ii) the so-called ``quantum'' 
part, that is, the velocity $\vec{\Vbf}$ of the motion {\it in} the 
CM frame (namely, the internal ``spin motion'' or {\it zitterbewegung}). \
By inserting such a complete, composite expression of the velocity into 
the kinetic energy term of the non-relativistic classical (i.e., newtonian) 
lagrangian, we straightforwardly get the appearance of the so-called 
``quantum potential" associated, as it is known, with the Madelung fluid.  
This result carries further evidence that {\it the quantum behaviour of 
micro-systems 
can be a direct consequence of the fundamental existence of spin}. \ \ 
In the second part (Sections 3 and 4), we fix our attention on the 
total velocity $\vec{\vbf} = \vec{\wbf} + \vec{\Vbf}$, it being now 
necessary to pass  to relativistic 
(classical) physics; and we show that the proper time entering the 
definition of the four-velocity $v^\mu$ for spinning particles has to be 
the proper time $\tau$ of the CM frame. \ Inserting the correct Lorentz 
factor into the
definition of $v^\mu$ leads to completely new kinematical properties for
$v^2$. \ The important constraint  \ $p_{\mu}v^{\mu} = m \,$, \  
identically true for scalar particles, but just assumed a priori in all
previous spinning particle theories, is herein {\it derived} in a 
self-consistent way.
\keywords {ICTE-1995, \LaTeX, spin, zitterbewegung, ``extended-like" 
particles, Madelung fluid, hydrodynamical formulation, 
Schroedinger theory, ``quantum" potential, Pauli current, K\"onig theorem, 
velocity for relativistic spinning particles, Barut--Zanghi theory,
Dirac current, Gordon decomposition, tensor algebra, Asim O. Barut.}
\end{abstract}

\

\hfill{{\em ``If a spinning particle is not quite a point particle, nor }}

\hfill{{\em a solid three dimensional top, what can it be?"}}

\rightline{Asim O. Barut \qquad \qquad}

\

\section{Madelung fluid: A variational approach}

\h The lagrangian for a non-relativistic scalar particle may be assumed to be:
\bb
\Le = \frac{i\hbar}{2}(\psi^{\star}\pa_t\psi - (\pa_t\psi^{\star})\psi)
 - \frac{\hbar^2}{2m}{\nablabf}\psi^{\star} \cdot {\nablabf}\psi
 - U\psi^{\star}\psi
\ee
where $U$ is the external potential energy and the other symbols have
the usual meaning. \ It is known that, by taking
the variations of $\Le$ with respect to $\psi, \; \psi^{\star}$, one can get
the Schroedinger equations for $\psi^{\star}$ and $\psi$, respectively.

\h By contrast, since a generic
scalar wavefunction $\psi \in$  I$\!\!\!$C can be written as
\bb
\psi = \sqrt{\rho} \; {\rm exp} [i \varphi / \hbar] \ ,
\ee
with $\rho,\varphi \in$ I$\!$R, we take the variations of
\bb
\Le = -\left[\pa_{t}\varphi + \frac{1}{2m}({\nablabf}\varphi)^2
 + \frac{\hbar^2}{8m}\left(\frac{{\nablabf}\rho}{\rho}\right)^2 + U\right]\rho
\ee
with respect to (w.r.t.) $\rho$ and $\varphi$. \ We then obtain[1-3]
the two equations for the so-called {\it Madelung fluid\/}$[4]$
(which, taken together, are equivalent to the Schroedinger equation):
\bb
\pa_{t}\varphi + \frac{1}{2m}({\nablabf}\varphi)^2
 + \frac{\hbar^2}{4m}\left[\frac{1}{2}\left(\frac{{\nablabf}\rho}{\rho}\right)^2
 - \frac{\triangle \rho}{\rho}\right] + U = 0 
\ee
and
\bb
\pa_t \rho + {\nablabf}\cdot (\rho {\nablabf}\varphi /m)
 = 0 \ ,
\ee
which are the Hamilton--Jacobi and the continuity
equation for the ``quantum fluid", respectively; \ where 
\bb
\frac{\hbar^2}{4m}\left[\frac{1}{2}\left(\frac{{\nablabf}\rho}{\rho}\right)^2
 - \frac{\triangle \rho}{\rho}\right] \equiv -\frac{\hbar^2}{2m}\frac{\triangle
|\psi|}{|\psi|}
\ee
is often called the ``quantum potential".  Such a potential  
derives from the last-but-one term in the r.h.s. of eq.(3), that is to say, 
from the (single) ``non-classical term" 
\bb
\frac{\hbar^2}{8m}\left(\frac{{\nablabf}\rho}{\rho}\right)^2   
\ee
entering our lagrangian $\Le$. 

\h Notice that we got the present {\it hydrodynamical reformulation} of the 
Schroedin-ger theory directly from a variational approach.$[3]$ 
This procedure, as we are going to see, offers us a physical interpretation 
of the non-classical terms appearing in eqs.(3) or (4). \ \ On the 
contrary, eqs.(4-5) are ordinarily obtained by inserting relation (2) into
the Schroedinger equation, and then separating the real and the imaginary 
part: a rather formal procedure, that does not shed light on the underlying
physics.
  
\h Let us recall that an early physical interpretation of the so-called
``quantum" potential, that is to say, of term (6) was forwarded
by de Broglie's pilot--wave theory$[5]$; in the
fifties, Bohm$[6]$ revisited and completed de Broglie's approach in
a systematic way [and, sometimes, Bohm's theoretical formalism
is referred to as the ``Bohm formulation of quantum mechanics'', alternative
and complementary to Heisenberg's (matrices and Hilbert spaces),
Schroedinger's (wave-functions), and Feynman's (path integrals) theory]. \
From Bohm's up to our days, several conjectures
about the origin of that mysterious potential have been put forth,
by postulating ``subquantal'' forces, the presence of an ether, and so on. \ 
 \ Well-known are also the derivations of the Madelung fluid within the
stochastic mechanics framework:$[7,2]$ in those theories, the origin of the 
non-classical term (6) appears as substantially {\it kinematical}. \ In the 
non-markovian approaches,$[2]$ for instance, after having assumed the 
existence of the so-called
zitterbewegung, a spinning particle appears as an
extended-like object, while the ``quantum" potential is tentatively related
to an internal motion. 

\h But we do not need following any stochastic approach, even if our 
phylosophical starting point is the {\it recognition} of the 
existence[8-12] of a {\it zitterbewegung} (zbw) or diffusive or
{\it internal\/} motion [i.e., of a motion observed {\bf in} the center-of-mass
(CM) frame, which is the one where $\imp = 0$ by definition], besides of
the [external, or drift, or translational, or convective] motion {\bf of} 
the CM. \ In fact, the existence of such an internal motion is denounced,
besides by the mere presence of spin, by the
remarkable fact that in the standard Dirac theory the
particle impulse $\imp$ is in general {\it not} parallel to the velocity: \
$\vbf \neq \imp /m$; \ moreover, while \ $[\wide{\imp},
\wide{H}]=0$ \ so that $\imp$ is a conserved quantity, quantity
$\vbf$ is {\it not} a constant of the motion: \ $[\wide{\vbf}, \wide{H}]\neq
0 \ \ (\wide{\vbf} \equiv \albf \equiv \ga^0\gabf$ being the usual vector 
matrix of Dirac theory). \
 Let us explicitly notice, moreover, that for dealing with the zbw it is 
highly convenient$[10,12]$ to split the motion
variables as follows (the dot meaning derivation with respect to time):
\bb
\xbf = \xibf + \Xbf \; ; \ \ \ \dot{\xbf} \equiv \vbf = \wbf + \Vbf \ ,
\ee
where $\xibf$ and $\wbf \equiv \dot{\xibf}$ describe
the motion of the CM in the chosen reference
frame, whilst $\Xbf$ and $\Vbf \equiv \dot{\Xbf}$ describe the
internal motion referred to the CM frame (CMF). \ [Notice that what is called
the ``diffusion velocity" $\vbf_{\rm dif}$ in the stochastic approaches is
nothing but our $\Vbf$]. \ From a dynamical point 
of view, the conserved electric current is associated  with the helical
trajectories[8-10] of the electric charge (i.e., with $\xbf$ and $\vbf \equiv 
\dot{\xbf}$), whilst 
the center of the particle coulombian field is associated with the 
geometrical center of such trajectories (i.e., with $\xibf$ and $\wbf \equiv
\dot{\xibf} = \imp /m$).

\h Going back to lagrangian (3), it is now possible
to attempt an interpretation$[3]$ of the non-classical term
${\frac{\hbar^2}{8m}} ({\nablabf\rho}/{\rho})^2$ appearing therein.
So, the first term in the r.h.s. of eq.(3) represents, apart from the sign,
the total energy
\bb
\pa_t \varphi = - E \ ;
\ee
whereas the second term is recognized to be
the kinetic energy  $\, \imp^{2}/2m$ \ {\it of} the CM, if one assumes that
\bb
\imp = - \nablabf \varphi.
\ee
The third term, that gives origin to the quantum potential, will be 
shown below to be interpretable as the kinetic energy {\it in} the CMF, 
that is, the internal energy due to the zbw motion. \ It will be soon
realized, therefore, that in lagrangian (3) the sum
of the two kinetic energy terms, ${\imp^2}/2m$ and
$\frac{1}{2}m\Vbf^2$, is nothing but {\it a mere application
of the K\"onig theorem}. \  We are not going to exploit, as often done,
the arrival point, i.e. the Schroedinger equation; by contrast,
we are going to exploit a non-relativistic (NR) analogue of the
Gordon decomposition$[13]$ of the Dirac current: namely, a suitable 
decomposition of the {\it Pauli current}.$[14]$ \ In so doing, we shall 
meet an interesting relation between  zbw and spin.\\

\section{The ``quantum" potential as a mere consequence of spin and zbw}

\h During the last thirty years Hestenes$[15]$ did
sistematically employ the Clifford algebras language in the description of
the geometrical, kinematical and hydrodynamical (i.e., {\it field\/})
properties of spinning particles, both in relativistic and NR physics, i.e.,
both for Dirac theory and for Schroedinger--Pauli theory. In the 
small-velocity limit of the Dirac equation, or directly from Pauli 
equation, Hestenes got the following decomposition of the particle velocity:
\bb
\vbf = \frac{\imp - e\Abf}{m} + \frac{\nablabf \wedge (\rho\sbf)}{m\rho}
\ee
where the light speed $c$ is assumed equal to 1, quantity $e$ is the 
electric charge, $\Abf$ is the external electromagnetic vector potential, 
$\sbf$ is the
{\it spin vector\/} $\sbf \equiv \rho^{-1}\psi^{\age}\wide{\sbf}\psi$,
and  $\wide{\sbf}$ is the spin operator usually represented in terms of
Pauli matrices as
\bb
\wide{\sbf} \equiv \frac{\hbar}{2}(\sig_{x}; \; \sig_{y}; \; \sig_{z}).
\ee
[Hereafter, every quantity is a {\it local}
or {\it field} quantity: ${\vbf} \equiv {\vbf} ({\xbf} ;t); \
{\imp} \equiv {\imp} ({\xbf} ;t); \
{\sbf} \equiv {\sbf} ({\xbf} ;t)$; and so on]. \ \ As a consequence, 
the internal (zbw) velocity reads:
\bb
\Vbf \equiv \frac{\nablabf\wedge (\rho\sbf)}{m\rho}.
\ee
\h Let us repeat the previous derivation ---by making now recourse to the  
ordinary tensor language--- {\it from} the familiar expression of the Pauli
current$[14]$ (i.e., from the Gordon decomposition of the Dirac
current in the NR limit):
\bb
\jbf = \frac{i\hbar}{2m}[(\nablabf \psi^{\age})\psi - \psi^{\age} \nablabf \psi] -
\frac{e\Abf}{m}\psi^{\age}\psi +
\frac{1}{m}\nablabf \wedge (\psi^{\age} \wide{\sbf}\psi) \ .
\ee
A spinning NR particle can be simply factorized into
\bb
\psi \equiv \sqrt{\rho}\> \Phi \ ,
\ee
$\Phi$ being a Pauli 2-component spinor, which
has to obey the normalization constraint
\[
\Phi^{\age}\Phi = 1
\]
if we want to have  $|\psi|^2 = \rho$.   

\h By definition \ $\rho\sbf \equiv \psi^{\age}\wide{\sbf}\psi \equiv
\rho\,\Phi^{\age}\wide{\sbf}\Phi$; \ therefore, introducing the
factorization $\psi \equiv \sqrt{\rho}\> \Phi$  
into the above expression (14) for the Pauli current, one just obtains:$[3]$
\bb
\jbf \; \equiv \; \rho \, \vbf \; = \; \rho \: \frac{\imp - e\Abf}{m} + 
\frac{\nablabf\wedge(\rho\sbf)}{m}
\ee
which is nothing but Hestenes' decomposition (11) of $\vbf$.

\h The Schroedinger subcase [i.e., the case in which the vector spin field
$\sbf = \sbf(\xbf, t)$ is constant in time and uniform in space] corresponds
to {\it spin eigenstates}; so that we need now a wave-function factorizable
into the product of a ``non-spin'' part $\sqrt{\rho}e^{i\varphi}$ 
({\it scalar\/}) and of a ``{\it spin}'' {\it part} $\chi$ (Pauli spinor):
\bb
\psi \equiv \sqrt{\rho}\, e^{i\frac{\varphi}{\hbar}}\chi \ ,
\ee
$\chi$ being {\it constant in time and space}. \ Therefore, when $\sbf$ has 
no precession (and no external field is present: $\Abf = 0$), we have \
$\sbf \equiv \chi^{\age}\wide{\sbf}\chi =$ constant, and
\bb
\Vbf = \frac{\nablabf \rho \wedge \sbf}{m\rho} \ne 0 \ . \ \ \ \ \ \ \ \
{\rm{(Schroedinger \ case)}}
\ee
One can notice that, {\it even in the Schroedinger theoretical framework, the 
zbw does not vanish}, except for plane waves, i.e., for the non-physical case
of $\imp$-eigenfunctions, when not only $\sbf$, but also $\rho$ is constant 
and uniform, so that $\nablabf \rho = 0$. \ [Notice also that the continuity 
equation (6), \ $\pa_t \rho + \nablabf\cdot (\rho\imp /m) = 0$, \  can be still 
rewritten in the ordinary way \ $\pa_t \rho + \nablabf\cdot (\rho\vbf) = 0$. \
 In 
fact, quantity $\nablabf\cdot\Vbf\equiv\nablabf\cdot \left( \nablabf\wedge 
(\rho\sbf) \right)$ is identically zero, it being the divergence of a rotor,
so that $\nablabf\cdot (\imp /m) = \nablabf\cdot\vbf$].

\h But let us go on. \ We may now write
\bb
\Vbf^2 = \left(\frac{\nablabf\rho\wedge\sbf}{m\rho}\right)^2 =
\frac{(\nablabf\rho)^2\sbf^2 - (\nablabf\rho\cdot\sbf)^2}{(m\rho)^2}
\ee
since in general it holds
\bb
(\abf \wedge \bbf)^2 = \abf^2\bbf^2 - (\abf\cdot\bbf)^2 \ .
\ee
Let us now put into equation (19), for instance, Hestenes' constraint 
($\be$ being the Takabayasi angle$[16]$): \ $\nablabf \cdot (\rho\sbf) = 
- m\rho \sin \be \;$, \ which in the NR limit yields $\be = 0$ (``pure electron'')
or $\be = \pi$ (``pure positron''), so that: \ $\nablabf\cdot (\rho\sbf) = 0$ \
and in the Schroedinger case [$\sbf =$ constant; \ $\nablabf\cdot\sbf = 0$] 
becomes
\bb
\nablabf\rho\cdot\sbf = 0.
\ee
Then, eq.(19) does assume$[3]$ the important form
\bb
\Vbf^2 = \sbf^2\left(\frac{{\nablabf}\rho}{m\rho}\right)^2 \ ,
\ee
which does {\it finally} allow us to attribute to the so-called 
``non-classical"
term, eq.(7), of our lagrangian (3) the simple meaning of kinetic energy
of the internal (zbw) motion [i.e., of kinetic energy associated with the
internal (zbw) velocity $\Vbf$], provided that
\bb
\hbar \: = \: 2 \sbf \ .
\ee
In agreement with the already mentioned K\"onig theorem, such an internal 
kinetic energy does appear, in lagrangian (3), as correctly
added to the (external) kinetic energy ${\imp}^2 / 2m$ {\bf of} the CM \ 
[besides to the total energy (9) and the external potential energy $U$].

\h Vice-versa, if we assume (within a zbw philosophy) that $\Vbf$, eq.(22), 
is the velocity attached to the kinetic energy term (7), {\it then we
can deduce} eq.(23), i.e., we deduce that actually: 
\[
{|\sbf|} \: = \: {1 \over 2} \, \hbar \ .
\]

\h Let us mention, by the way, that in the stochastic approaches the 
(``non-classical") stochastic, diffusion velocity is $\Vbf \equiv 
\vbf_{\rm dif} = \nu \, ({{\nablabf \rho} / \rho})$, quantity $\nu$ being 
the diffusion coefficient of the ``quantum" medium. \ In those approaches,
however, one has to {\it postulate} that \ $\nu \equiv \hbar / 2m$. \ In our
approach, on the contrary, if we just adopted for a moment the stochastic 
language, by comparison of our eqs.(7), (22) and (23)  we would immediately
{\it deduce} that $\nu = \hbar / 2m$ and therefore the interesting relation
\bb
\nu = \frac{|\sbf|}{m} \ .
\ee

\h Let us explicitly remark that, because of eq.(22), in the Madelung fluid 
equation (and therefore in the
Schroedinger equation) quantity $\hbar$ is naturally replaced by 
$2|\sbf|$, the presence itself of the former quantity being no longer needed; in a way, we
might say that it is more appropriate to write $\hbar = 2|\sbf|$, rather
than\break 
$|\sbf| = \hbar /2$ \ldots!

\h Let us {\bf conclude} the {\it first part} of the present contribution 
by stressing 
the following. \ We first achieved a non-relativistic, Gordon-like
decomposition of the field velocity within the ordinary
tensorial language. \ Secondly, we derived the ``quantum" potential
(without the postulates and 
assumptions of stochastic quantum mechanics) by simply
relating the ``non-classical" energy term to zbw and spin. \ Such results
carry further evidence that {\bf the quantum behaviour
of micro-systems may be a direct consequence of the existence of spin.}
In fact, when $\sbf = 0$, the quantum
potential does vanish in the Hamilton--Jacobi equation, which then becomes 
a totally {\it classical} and newtonian equation. \ We have also seen that
quantity $\hbar$ itself enters the Schroedinger equation owing to the
presence of spin. \ We are easily induced to conjecture that no scalar
{\it quantum} particles exist that are really elementary; \ but that scalar 
particles are always constituted by spinning objects endowed with 
zbw.

\section{About the kinematics of spinning particles}

\h In the first part of this paper, we addressed ourselves to spin, zbw and
Madelung fluid in (non-relativistic) physics. \ The previous analysis led
us, in particular, to fix our attention on the internal velocity $\Vbf$ 
of the spinning particle, besides on its external velocity $\wbf = \imp / m$. 
 \ In the {\it second part} of this article, we want to fix our attention
on the {\it total} velocity $\vbf = \wbf + \Vbf$. It is now essential to
allow $\wbf$ assume any value, and therefore to pass to {\it relativistic}
physics. In what follows our considerations will be essentially 
{\it classical}, while the quantum side of these last Sections will be 
studied in the next contribution to this Volume.$[17]$

\h Before going on, let us make a brief digression by recalling that, since 
the works by Compton,$[8]$  Uhlenbeck and Goudsmith,$[18]$   
Frenkel,$[18]$ and Schr\"odinger$[9]$
till the present times, many classical theories ---often quite different 
among themselves from a physical
and formal point of view--- have been advanced for spinning particles [for
simplicity, we often write ``spinning particle'' or just ``electron" instead  of the more
pertinent expression ``spin-$1 \over 2$ particle'']. \ Following
Bunge,$[19]$ they can be divided into
three classes:

\h I) strictly {\it point-like} particle models

\h II) actual extended--type particle models 
(``spheres'', ``tops'', ``gyroscopes'', and so on)

\h III)  mixed models for ``extended--like" particles, in which 
the center of the {\it point-like} charge $\cal Q$ results to be spatially
distinct from the particle center-of-mass (CM).

\h Notice that in the theoretical approaches of type III ---which, being 
in the middle between classes I and II, could answer the dilemma posed 
by Barut at the top of this paper---
the motion of $\cal Q$ does not coincide with the motion of the particle CM.
This peculiar feature was found to be an actual characteristic[20-22,15,11,10] 
(just called, as we know, the zbw motion) of spinning particles kinematics. \
The type III models, therefore, are a priori convenient for describing zbw,
spin and intrinsec magnetic moment of the electron,
while these properties are hardly predicted by making recourse to the
point-like--particle theories of class I. \  The theories of type III, 
moreover, are consistent[8-12] with the ordinary quantum theory of the 
electron: see below. \  The ``extended--like" electron 
models of class III are at present after fashion also because of their 
possible generalizations to include supersymmetry and superstrings.$[10b]$   
 \ At last, the ``mixed'' models help bypassing
the obvious non-locality problems involved by a relativistic
covariant picture for extended--type (in particular {\it rigid\/}) objects 
of class II. \ Quite differently, the extended--like (class III) electron is  
non-rigid and consequently variable in its ``shape ''and in its
characteristic ``size'', depending on the considered dynamical 
situation. This is a priori consistent with  the appearance
in the literature of many different ``radii of the electron'' [for 
instance, in his book,[23] McGregor lists at page 5 seven typical electron 
radii, from the Compton to the ``classical'' and to the ``magnetic'' 
radius]. \  Because of all these reasons, therefore, the spinning particle 
we shall have in mind in the next Section is to be described by 
class III theories.

\h We have here to rephrase some of the previous considerations in terms of  
Minkowsky (four-dimensional) vectors. For instance, let us recall again
that in the ordinary  Dirac theory the particle four-impulse $p^\mu$ is 
in general {\it not} parallel to the four-velocity: \ $v^\mu \neq p^\mu/m$. 
 \ Before all, let us repeat that, in order to describe the zbw, 
in all type III theories it is very convenient[10-12] to split the motion
variables as follows (the dot meaning {\it now} derivation {\it with respect 
to the proper time} $\tau$):
\bb
x^\mu \equiv \xi^\mu + X^\mu \; ; \ \ \ \dot{x}^\mu \equiv v^\mu = w^\mu + V^\mu \ ,
\ee
where $\xi^\mu$ and $w^\mu\equiv \dot{\xi}^\mu$ describe as before
the external motion, i.e. the motion of the CM,
whilst $X^\mu$ and $V^\mu \equiv \dot{X}^\mu$ describe the internal motion. \
From an electrodynamical point of view, as we know, the conserved 
electric current is associated with the
trajectories of $\cal Q$ (i.e., with $x^\mu$), whilst the center of the
particle Coulomb field ---obtained,$[22]$  e.g., through a time average  
over the field generated by the quickly oscillating charge---
is associated with the CM  (i.e., with $w^\mu$; and then, for free particles,
with the geometric center of the internal motion). \ 
In such a way, it is $\cal Q$ which follows {\it the (total) motion}, whilst
the CM follows the {\it mean motion} only. \ It is worthwile also to notice
that the classical extended--like electron of type III
is totally consistent with the standard Dirac theory; in fact, the above
decomposition for the total motion is the classical analogue of two 
well-known
quantum-mechanical procedures: i.e., of the {\it Gordon decomposition}
of the Dirac current, and the (operatorial) {\it decomposition of the 
Dirac position operator}
proposed by Schr\"odinger in his pioneering works.$[9]$
We shall come back to these points below.

\h The well-known Gordon decomposition of the Dirac current reads$[13]$ 
(hereafter we shall choose units such that numerically $c=1$):
\bb
\psib\ga^\mu\psi = \frac{1}{2m}\,[\psib\po^\mu\psi - (\po^\mu\psib)\psi]
- \frac{i}{m}\po_\nu\,(\psib S^{\mu\nu}\psi)\; ,
\ee
$\psib$ being the ``adjoint'' spinor of $\psi$; \ quantity 
$\po^\mu \equiv i\pa^\mu$
the 4-dimensional impulse operator; and $S^{\mu\nu} \equiv
\frac{i}{4}\,(\ga^\mu\ga^\nu - \ga^\nu\ga^\mu)$
the spin-tensor operator.
 \ The ordinary interpretation of eq.(26) is in total analogy 
with the decomposition given in eq.(25).
The first term in the r.h.s. results to be associated with
the translational motion of the CM ({\it scalar} part of the current, 
corrisponding to
the traditional Klein--Gordon current).
The second term in the r.h.s. results, instead, directly connected with 
the existence of spin, and describes the zbw motion.

\h In the abovequoted papers, Schr\"odinger started from the Heisenberg equation
for the time evolution of the acceleration operator in Dirac theory [$\vbf 
\equiv \albf$]
\bb
\abf \equiv \frac{\drm\vbf}{\drm t} \ug {i\over\hbar}\,[H, \vbf] \ug
{2i \over \hbar} \, (H\vbf - \imp)\; ,
\ee
where $H$ is equal as usual to $\vbf$$\cdot$$\imp + \bt\,m$. \  
Integrating once this operator equation over time, after some
algebra one can obtain:
\bb
\vbf \ug H^{-1}\imp - {i \over 2} \hbar \, H^{-1}\abf\; ,
\ee
and, integrating it a second time, one obtains$[14]$ just the spatial part 
of the decomposition:
\bb
\xbf \equiv \xibf + \Xbf
\ee
where (still in the operator formalism) it is
\bb
\xibf = \rbf + H^{-1}\imp t \; ,
\ee
related to the motion of the CM, \ and
\bb
\Xbf = {i \over 2} \hbar \, \etbf H^{-1} \; ,  \qquad  (\etbf \equiv \vbf - 
H^{-1}\imp) \; ,
\ee
related to the zbw motion.

\section{New kinematical properties of the ``extended--like" particles}

\h We want now to analyze the formal and conceptual properties of a {\it new
definition} for the 4-velocity of our extended--like electron.
Such a new definition has been first adopted ---but without any emphasis---
in the papers by Barut {\it et al.}
dealing with a successful model for the relativistic classical 
electron.$[10a,12]$  \ Let us consider the following. 
At variance with the procedures followed in the literature
from Schr\"odinger's till our days, we have to make recourse {\it not to 
the proper time of the
charge $\cal Q$, but rather to the proper time of the
center-of-mass}, i.e. to the time of the CMF.$^{\; \# 1}$ \ \ 
\footnotetext{$^{\; \# 1}$ Let us recall once more that the CMF is the frame 
in which the kinetic impulse vanishes identically, $\vec{\imp} = 0$.  For spinning
particles, in general, it is {\it not} the ``rest'' frame, since the
velocity $\vec{\vbf}$ is not necessarily zero in the CMF.}
As a consequence, quantity $\tau$ in the denominator of the 4-velocity 
definition, $v^\mu \equiv \drm x^\mu/ \drm \tau$, has to be
the {\it latter} proper time. \
Up to now ---with the exception of the above-mentioned papers by 
Barut {\it et al.}--- in all
theoretical frameworks the Lorentz factor has been assumed to be equal to
$\sqrt{1 - {\bf v}^2}$. \ On the 
contrary, into the Lorentz factor it has to enter 
${\bf w}^2$ instead of ${\bf v}^2$, \  
quantity  ${\bf w} \equiv {\bf p}/{p^0}$ being the 3-velocity of the CM 
with respect to the chosen frame [$p^0 \equiv {\cal E}$ is the energy]. \ 
By adopting the correct Lorentz factor, all the formulae containing
it are to be rewritten, and {\it they get a new physical meaning}. \ 
In particular, we shall show below that the new definition does actually 
{\it imply\/}$^{\# 2}$ the important
constraint, which ---holding identically for scalar particles--- is often 
just {\it assumed} for spinning particles:
\footnotetext{$^{\# 2}$ For all plane wave solutions $\psi$ of the Dirac
equation, we have (labelling by $< >$ the corresponding
{\it local mean value} or {\it field density}): \
$p_{\mu}< \wide{v}^\mu > \equiv p_{\mu}\psi^{\dag} \wide{v}^\mu \psi \equiv
p_{\mu}\psi^{\dag} \ga^0 \ga^\mu \psi \equiv p_{\mu}\ov{\psi} \ga^\mu \psi
= m$.}
\[
p_{\mu}v^{\mu} = m   \ ,
\]
where $m$ is the {\it physical} rest mass of the particle (and not an
ad hoc mass-like quantity $M$).$^{\# 3}$\\
\footnotetext{$^{\# 3}$ Let make just an example, recalling that
Pa\v{v}si\v{c}$[10b]$ derived, from a lagrangian containing an 
{\it extrinsic curvature}, the classical equation of the motion
for a rigid $n$-dimensional world-sheet in a curved background spacetime. \
Classical world-sheets describe {\it membranes} for
$n\geq3$, strings for $n=2$, and point particles for $n=1$. \
For the special case $n=1$, he found nothing but the traditional 
Papapetrou equation for a classical spinning particle; also,
by ``quantization'' of the classical theory, he actually  derived
the Dirac equation.  In ref.$[10b]$, however, $M$ is not the observed 
electron mass $m$:
and the relation between the two masses reads: $m= M+\mu H^2$, quantity
$\mu$ being the so-called {\it string rigidity}, while $H$ is the second
covariant derivative on the world-sheet.}

\h Our choice of the proper time $\tau$ may be supported by the
following considerations:

\h (i) The {\it light-like} zbw ---when the speed of $\cal Q$ is constant 
and equal to
the speed of light in vacuum--- is certainly the preferred one 
(among all the ``a priori'' possible internal motions) in the literature, 
and to many authors it appears 
the most adequate for a meaningful classical picture
of the electron. \ In some special theoretical approaches,
the light speed is even regarded as the {\it quantum-mechanical} typical 
speed for the zbw.
In fact, the Heisenberg principle in the relativistic domain$[14]$ 
implies (not controllable) particle--antiparticle pair creations
when the (CMF) observation involves space distances of the order of a Compton 
wavelenght. So that $\hbar / m$ is assumed to be the characteristic
``orbital" radius and $2m / \hbar^2$ the (CMF) angular frequency
of the zbw ---as first noticed by Schr\"odinger;---  \ and the orbital motion of  
$\cal Q$ is expected to be {\it light-like}.  \ \
Now, if the charge $\cal Q$ travels at the light speed, {\it the
proper time
of $\cal Q$ does not exist}; while the proper time of
the CM (which travels at sub-luminal speeds) does exist. \
Adopting as time the proper time of $\cal Q$, as often done in the past 
literature, automatically {\it excluded} a
light-like zbw. In our approach, by contrast, such zbw motions are 
not excluded. \
Analogous considerations may hold for {\it Super-luminal} zbw speeds, 
without too much problem, since the CM (which carries the
energy-impulse and the ``signal") is always endowed with a subluminal motion;

\h (ii) The indipendence between the center-of-charge and
the center-of-mass motion becomes evident by our definition. As a
consequence {\it the non-relativistic limit can be formulated by us in a 
correct, and univocal, way}. Namely, by assuming the correct Lorentz 
factor, one can immediately see
that the zitterbewegung can go on being a relativistic (in particular, 
light-like) motion
even in the non-relativistic approximation: i.e., when $\imp \longr 0$
(this is perhaps connected with the non-vanishing of spin in 
the non-relativistic limit). \ In fact, in the non-relativistic limit, 
we have to take
\[
\wbf^2 \ll 1 \ ,
\]
and {\it not} necessarily
\[
\vbf^2 \ll 1
\]
as usually assumed in the past literature;

\h (iii) The definition for the 4-velocity that we are going to propose [see
eq.(33) in the following] does agree with the natural
``classical limit" of the Dirac current. Actually, it was used in those 
models which (like Barut {\it et al.\/}'s) define velocity
{\it even at the classical level}
as the bilinear combination $\psib\ga^\mu\psi$, 
via a direct introdution of {\it classical} spinors $\psi$. \
By the new definition, we shall be able to write the translational
term as $p^{\mu}/m$, with the {\it physical} mass in the denominator, exactly
as in the Gordon decomposition, eq.(26). \ Quite differently, in all the
theories adopting as time the proper time of $\cal Q$, it appears in the 
denominator an ad-hoc {\it variable} mass $M$, which depends on the internal 
zbw speed $V$ (see below);

\h (iv) The choice of the CM proper time constitutes a natural extension of
the ordinary procedure for relativistic scalar particles. In fact,
for spinless particles in relativity the 4-velocity is known to be univocally
defined as the derivative of 4-position with respect to the CMF
proper time (which is the only one available in that case).\\

\h The most valuable reason in support of our definition turns out to be
the circumstance that the {\it old definition}
\bb
v_{\rm std}^\mu = (1/ \sqrt{1- \vbf^2}; \;\; \vbf / \sqrt{1- \vbf^2})\;
\ee
seems to entail a mass varying with the internal zbw speed. 

\h But let us explicitate our {\it new} definition for $v^\mu$. \
The symbols which we are going to use possess the ordinary meaning; the
novelty$[24]$ is that now {\it the Lorentz factor $\drm\tau / \drm t$ will not be
equal to $\sqrt{1- \vbf^2}$, but instead to $\sqrt{1- \wbf^2}$ .} \
Thus we shall have:
\[
v^\mu\equiv \drm x^\mu/ \drm \tau \equiv (\drm t/ \drm \tau;\,
\drm \xbf / \drm \tau)
\equiv ({\frac{\drm t}{\drm \tau}};\; \frac{\drm \xbf}{\drm t} \;
\frac{\drm t}{\drm \tau})
\]
\bb
=(1/ \sqrt{1- \wbf^2}; \;\; \vbf / \sqrt{1- \wbf^2})\; . \;\;\;\;\;\; \quad
[\vbf \equiv \drm \xbf / \drm t] 
\ee
For $w^\mu$ we can write:
\[
w^\mu\equiv \drm \xi^\mu/ \drm \tau \equiv (\drm t/ \drm \tau;\,
\drm \xibf / \drm \tau)
\equiv ({\frac{\drm t}{\drm \tau}};\; \frac{\drm \xibf}{\drm t} \;
\frac{\drm t}{\drm \tau})
\]
\bb
=(1/ \sqrt{1- \wbf^2}; \;\; \wbf / \sqrt{1- \wbf^2})\; ;\;\;\;\;\;\; \quad
[\wbf \equiv \drm \xibf / \drm t]
\ee
and for the 4-impulse:
\bb
p^\mu \equiv mw^\mu = m (1/ \sqrt{1- \wbf^2}; \;\; \wbf / \sqrt{1- \wbf^2})
 \; .
\ee
[In presence of an external field such relations remain valid, 
provided that one
makes the ``minimal prescription": \ $p \longr p - eA$ \ (in the CMF we 
shall have
$\imp - e\Abf = 0$ and consequently $\wbf = 0$, as above)].

\h Let us now examine the resulting impulse--velocity scalar product, 
$p_\mu v^\mu$, which has to
be a Lorentz invariant, both with our $v$ and with the old $v_{\rm std}$. \ 
Quantity
$p \equiv (\vare;\; \imp)$ being the 4-impulse, and $M_1, M_2$ two relativistic
invariants, we may write:
\bb
p_\mu v^\mu \equiv M_1 \equiv \frac{\vare - \imp\cdot\vbf}{\sqrt{1- \wbf^2}}\; ,
\ee
or, alternatively,
\bb
p_\mu v^\mu_{\rm std} \equiv M_2 \equiv \frac{\vare - \imp\cdot\vbf}
{\sqrt{1- \vbf^2}}\; .
\ee
If we refer ourselves to the CMF, we shall have \ $\imp_{{\rm CMF}} =
\wbf_{{\rm CMF}} = 0$ \ (but $\vbf_{{\rm CMF}}
\equiv \Vbf_{{\rm CMF}} \neq 0$), \ and then
\bb
M_1 = \vare_{{\rm CMF}}
\ee
in the first case; and
\bb
\vare_{{\rm CMF}} = M_2\,\sqrt{1- \Vbf_{{\rm CMF}}^2}
\ee
in the second case. \ \ So,
we see that the invariant $M_1$ is actually a {\it constant}, which ---being 
nothing but
the center-of-mass energy, $\vare_{{\rm CMF}}$--- can be identified, 
as we are going to prove, with
the physical mass $m$ of the particle. \ On the contrary, in the second
case (the standard one), the center-of-mass energy results to be 
{\it variable} with the internal motion.

\h Now, from eq.(35) we have
\[
p_{\mu}v^{\mu} \equiv m w_{\mu}v^{\mu}
\]
and, because of eqs.(33,34),
\bb
p_{\mu}v^{\mu} \equiv m (1 - {\wbf}{\vbf}) / (1- {\wbf}^2) \ .
\ee
Since $\wbf$ is a vector component of the total
3-velocity $\vbf$, due to eqs.(25), and moreover is the orthogonal 
{\it projection} of $\vbf$ along the $\imp$-direction, we can write
\[
\wbf \cdot \vbf = {\wbf}^2 \ ,
\]
which, introduced into eq.(40), yields$[24]$ the important relation:
\bb
m = p_{\mu}v^{\mu} \ .
\ee

\h Quite differently, by use of the wrong Lorentz factor, we would
have got
\[
v^\mu = (1/ \sqrt{1- \vbf^2}; \;\; \vbf / \sqrt{1 - \vbf^2})
\]
and consequently
\[
p_{\mu}v^{\mu} \equiv m (1 - \wbf\vbf) / \sqrt{(1 - \wbf^2)(1 - \vbf^2)}
\]
\[
= m \sqrt{1 - \wbf^2} / \sqrt{1 - \vbf^2} \neq m \ .
\]

\h By recourse to the correct Lorentz factor, therefore, we succeeded in 
showing that the noticeable constraint \ $m \ug p_{\mu}v^{\mu}$, \
trivially valid for scalar particles, does hold for spinning particles too.
 \ Such a relation, eq.(41), would be very useful also for a hamiltonian 
formulation of the electron theory.$[12]$\\

\h Finally, we want to show that the ordinary kinematical 
properties of the Lorentz invariant $v^2 \equiv v_{\mu}v^{\mu}$ do {\it not} 
hold any longer 
in the case of spinning particles, endowed with zitterbewegung. In fact, 
it is easy to prove
that the ordinary constraint for scalar relativistic
particles ---quantity $v^2$ constant in time and equal to~$1$---
does {\it not} hold for spinning particles endowed with zbw.
 \ Namely, if we choose as reference frame the CMF, in which  $\wbf = 0$,
we have [cf. definition (33)]:
\bb
v^\mu_{\rm CMF} \equiv (1; \Vbf_{{\rm CMF}}) \ ,
\ee
{\it wherefrom}, it being
\bb
v^2_{\rm CMF} \equiv 1 - \Vbf^2_{{\rm CMF}}\; ,
\ee
one can deduce$[24]$ the following {\it new constraints}:
\[
0 < \Vbf_{{\rm CMF}}^2 (\tau) < 1  \;\;\; \Lef \;\;\; 
0 < v^2_{{\rm CMF}} (\tau) < 1  \;\;\;\;\;\; \quad\mbox{(``time-like")}
\]
\bb
\Vbf^2_{{\rm CMF}} (\tau) = 1  \quad \Lef \quad 
v^2_{{\rm CMF}}(\tau)=0  \;\;\;\;\; \ \ \ \ \ \ \ \;
\quad\mbox{ \ \ (``light-like")}
\ee
\[
\Vbf^2_{{\rm CMF}} (\tau) > 1  \quad \Lef \quad 
v^2_{{\rm CMF}} (\tau) < 0  \ . \;\;\;\;\; \ \ \ \ \ \ \ \ 
\quad\mbox{ \ \ (``space-like")} 
\]
Since the square of the total 4-velocity is invariant and in 
particular it is $v^2_{{\rm CMF}} = 
v^2$, these new constraints for $v^2$ will be valid in any frame:
\[
0 < v^2 (\tau) < 1 \;\;\;\;\;\;\;\;\;\;\;\;
\quad\mbox{(``time-like'')}
\]
\bb
\ \ v^2 (\tau)=0  \ \ \;\;\;\;\;\;\;\;\;\;\;\;\;
\quad\mbox{(``light-like'')}
\ee
\[
\ \ \ \ v^2 (\tau) < 0 \ .  \;\;\;\;\;\;\;\;\;\;\;\;\;
\quad\mbox{(``space-like'')}
\]
Notice explicitly that the correct application of Special Relativity to a 
spinning particle led us, under our hypotheses, to obtain that $v^2 = 0$ 
in the light-like case, {\it but} $v^2 \, \ne \, +1$ in the
time-like case \ and \ $v^2 \, \ne \, -1$ in the space-like case.

\h Let us now examine the manifestation and consequences of such new 
constraints in a specific example: namely, in the
already mentioned theoretical model by Barut--Zanghi$[10a]$
which did implicitly adopt as time the proper time of the CMF. \
In this case, we get that in general it is $v^2\neq 1$. \ And in fact 
one obtains$[12]$ the remarkable relation:
\bb
v^2 = 1 - \frac{\ddot{v}_\mu v^\mu}{4 m^2} \ .
\ee
In particular,$[22]$ in the light-like case it is \ 
$\ddot{v}_\mu v^\mu = 4 m^2$ \ and therefore \ $v^2 = 0$.

\h Going back to eq.(43), notice that now quantity $v^2$ is no longer related
to the {\it external} speed $|\wbf|$ of the CM, but on the contrary to
the {\it internal} zitterbewegung speed $|\Vbf_{{\rm CMF}}|$. \ Notice at last
that, in general ---and at variance with the scalar case---
the value of $v^2$ {\it is not constant in time} any longer, but varies 
with $\tau$
(except when ${\Vbf^2_{{\rm CMF}}}$ itself is constant in time).

\acknowledgements {This work is dedicated to the memory of Asim O. Barut. \  
The authors wish to acknowledge stimulating discussions with R.J.S. 
Chisholm, H.E. Hern\'andez, J. Keller, Z. Oziewicz, W.A. Rodrigues and J. Vaz.
 \ For the kind cooperation, thanks are also due to G. Andronico, 
M. Baldo, A. Bonasera, M. Borrometi, A. Bugini, F. Catara, L. D'Amico, 
G. Dimartino, M. Di Toro, G. Giuffrida, C. Kiihl, L. Lo Monaco, 
G. Marchesini, R.L. Monaco, E.C. Oliveira, M. Pignanelli, G.M. Prosperi, 
R.M. Salesi, M. Sambataro, S. Sambataro, 
M. Scivoletto, R. Sgarlata, R. Turrisi, M.T. Vasconselos, J.R. Zeni, and 
particularly I. Arag\'on, C. Dipietro and J.P. dos Santos. \ One of the 
authors (ER) wishes to thank Prof. J. Keller and all the Organizers
for generous hospitality during this International Conference; and 
W.A. Rodrigues and FAPESP for a research grant.}

\end{document}